\documentstyle[12pt]{article}
\begin{document}
\begin{center}
{\bf
The fractal structure of the
universe: a new field theory approach,
 H.J. de Vega$^{(a)}$, N. S\'anchez$^{(b)}$ and  F. Combes$^{(b)}$}

(a) Laboratoire de Physique Th\'eorique et Hautes Energies,
Universit\'e Paris VI, UA 280, Tour 16, 1er \'etage, 4, Place Jussieu
75252 Paris, Cedex 05, France

(b) Observatoire de Paris, DEMIRM, 61, Avenue de l'Observatoire,
75014 Paris, France
\end{center}
\begin{abstract}
While the universe becomes more and more homogeneous at large scales,
statistical analysis of galaxy catalogs have revealed a fractal structure 
at small-scales ($\lambda < 100 h^{-1}$ Mpc), with a fractal dimension
$D=1.5-2$ (Sylos Labini et al 1996). We study the thermodynamics of a
self-gravitating system with the theory of critical phenomena and finite-size
scaling and show that gravity provides a dynamical mechanism to produce 
this fractal structure. We develop a field theoretical approach to
compute  the galaxy distribution, assuming them to be in
quasi-isothermal equilibrium. Only a limited, (although large), range
of scales is involved, between a short-distance cut-off below which other 
physics intervene, and a large-distance cut-off, where the thermodynamic 
equilibrium is not satisfied. The galaxy ensemble can be considered at
critical conditions, with large density fluctuations developping at any scale.
From the theory of critical phenomena, we derive the two independent critical 
exponents $\nu$ and $\eta$ and predict the fractal dimension $ D = 1/\nu $ to 
be either $1.585$ or $2$, depending on whether the long-range behaviour is 
governed by the Ising or the mean field fixed points, respectively. Both set 
of values are compatible with present observations. In addition, we predict the
scaling behaviour of the gravitational potential to be $ r^{- \frac12
(1 + \eta )} $. That is, $ r^{-0.5} $ for mean field or $ r^{- 0.519}
$ for the Ising fixed point. The theory allows to compute the three
and higher density correlators without any assumption or Ansatz. We find that 
the $N$-points density scales as $ r_1^{(N-1)(D-3)} $, when $ r_1 >> r_i, \; 
2\leq i \leq N $. There are no free parameters in this theory.
\end{abstract}


gravitation  ---
          galaxies           : clusters: general  ---
          galaxies           : general ---
          galaxies           : statistics  ---
          cosmology          : miscellaneous ---
          cosmology          : large-scale structure of universe

\section{Introduction}

One obvious feature of galaxy and cluster distributions in the sky is their
hierarchical property: galaxies gather in groups, that are embedded in
clusters, then in superclusters, and so on. (Shapley 1934, Abell 1958). 
Moreover, galaxies and clusters appear to obey scaling properties,
such as the power-law of the two point-correlation function:
$$
\xi(r) \propto r^{-\gamma}
$$
with the slope $\gamma$, the same for galaxies and clusters, of $\approx$ 1.7
(e.g. Peebles, 1980, 1993).  This scale-invariance has suggested very
early the idea of fractal models for the clustering hierachy of galaxies
(de Vaucouleurs 1960, 1970; Mandelbrot 1975). Since then, many authors have
shown that a fractal distribution indeed reproduces quite well the 
aspect of galaxy catalogs, for example by simulating a fractal and observing
it, as with a telescope (Scott, Shane \& Swanson, 1954; Soneira \& Peebles 
1978). Sometimes the analysis has been done in terms of a multi-fractal
medium (Balian \& Schaeffer 1989, Castagnoli \& Provenzale 1991, 
Martinez et al 1993, Dubrulle \& Lachieze-Rey 1994).

There is some ambiguity in the definition of the two-point correlation
function $\xi(r)$ above, since it depends on the assumed scale
beyond which the universe is homogeneous; indeed it includes a normalisation by
the average density of the universe, which, if the homogeneity scale is not 
reached, depends on the size of the galaxy sample.
Once $\xi(r)$ is defined, one can always determine a  length $r_0$ where 
$\xi(r_0)$ =1 (Davis \& Peebles 1983, Hamilton 1993).  For galaxies,
the most frequently reported value is $r_0 \approx 5 h^{-1}$ Mpc
(where $h = H_0$/100km s$^{-1}$Mpc$^{-1}$), but it has been shown to increase 
with the distance limits of galaxy catalogs (Davis et al 1988).
$r_0$ is called `correlation length' in the galaxy literature. 
[The notion of correlation length $\xi_0$ is usually different in physics,
where  $\xi_0$ characterizes the exponential decay of correlations $ (\sim 
e^{- r/ \xi_0} ) $. For power decaying correlations, it is said that the  
correlation length is infinite].

The same problem occurs for the two-point correlation function of
galaxy clusters; the corresponding $\xi(r)$ has the same power law 
as galaxies, their  length  $r_0$ has been reported to be about 
$r_0 \approx 25 h^{-1}$ Mpc, and their correlation amplitude is therefore
about 15 times higher than that of galaxies
(Postman, Geller \& Huchra 1986, Postman, Huchra \& Geller 1992).
The latter is difficult to understand, unless there is a considerable
difference between galaxies belonging to clusters and field galaxies (or
morphological segregation). The other obvious explanation is that
the normalizing average density of the universe was then chosen lower.

This statistical analysis of the galaxy catalogs has been criticized by
Pietronero (1987), Einasto (1989) and Coleman \& Pietronero (1992), 
who stress the unconfortable dependence of $\xi(r)$ and of the  length $r_0$
upon the finite size of the catalogs, and on the {\it a priori} assumed 
value of the large-scale homogeneity cut-off.  A way to circumvent these 
problems is to deal instead with the average density as a function of size 
(cf \S 2). It has been shown that the galaxy distribution behaves as a pure
self-similar fractal over scales up to $\approx 100 h^{-1}$ Mpc,
the deepest scale to which the data are statistically robust
(Sylos Labini et al 1996; Sylos Labini \& Pietronero 1996).
This is more consistent with the observation of contrasted large-scale
structures, such as superclusters, large voids or great walls of
galaxies of $\approx 200 h^{-1}$ Mpc (de Lapparent et al 1986, Geller 
\& Huchra 1989). After a proper statistical analysis of all available
catalogs (CfA, SSRS, IRAS, APM, LEDA, etc.. for galaxies, and Abell and
ACO for clusters) Pietronero et al (1997) state that the 
transition to homogeneity might not yet have been reached up to the deepest
scales probed until now. At best, this point is quite controversial,
and the large-scale homogeneity transition is not yet well known. 

Isotropy and homogeneity are expected at very large scales from the
Cosmological Principle (e.g. Peebles 1993). However, this does not imply
local or mid-scale homogeneity (e.g. Mandelbrot 1982, Sylos Labini 1994):
a fractal structure can be locally isotropic, but inhomogeneous.
The main observational evidence in favor of the Cosmological Principle
is the remarkable isotropy of the
cosmic background radiation (e.g. Smoot et al 1992), that provides information
about the Universe at the matter/radiation decoupling. There must therefore
exist a transition between the small-scale fractality to large-scale
homogeneity. This transition is certainly smooth, and might correspond to the
transition from linear perturbations to the non-linear gravitational collapse 
of structures. The present catalogs do not yet see the transition since
they do not look up sufficiently back in time. It can be noticed that
some recent surveys begin to see a different power-law
behavior at large scales ($\lambda \approx 200-400  h^{-1}$ Mpc, e.g. 
Lin et al 1996).

\medskip

There are several approaches to understand non-linear clustering, and therefore
the distribution of galaxies, in an infinite gravitating system. 
Numerical simulations have been widely used, in the hope to trace back from 
the observations the initial mass spectrum of fluctuations, and to test 
postulated cosmologies such as CDM and related variants (cf Ostriker 1993). 
This approach 
has not yet yielded definite results, especially since the physics of
the multiple-phase universe is not well known. Also numerical limitations
(restricted dynamical range due to the softening and limited volume) have
often masked the expected self-similar behavior (Colombi et al 1996).
 A second approach, which 
should work essentially in the linear  (or weakly non-linear) regime, 
is to solve the BBGKY hierarchy
through closure assumptions (Davis \& Peebles 1977; Balian \& Schaeffer 1989).
The main assumption is that the $N$-points 
correlation functions are
scale-invariant and behave as power-laws like is observed for the few-body
correlation functions. Crucial to this approach is the determination of
the void probability, which is a series expansion of the 
$N$-points correlation 
functions (White 1979). The hierachical solutions found in this frame
agree well with the simulations, and with the fractal structure of the universe
at small-scales (Balian \& Schaeffer 1988). 
A third approach is the thermodynamics of gravitating systems, developped
by Saslaw \& Hamilton (1984), which assumes quasi thermodynamic
equilibrium. The latter is justified at the small-scales of non-linear
clustering, since the expansion time-scale is slow with respect to local
relaxation times. Indeed the main effect of expansion is to subtract 
the mean gravitational field, which is negligible for structures of
mean densities several orders of magnitude above average.
  The predictions of the thermodynamical theory have been successfully
compared with N-body simulations (Itoh et al 1993), but a special physical
parameter (the ratio of gravitational correlation energy to thermal energy) 
had to be adjusted for a better fit (Bouchet et al 1991, Sheth \& Saslaw 1996,
Saslaw \& Fang 1996). 

We present in this article a new approach based on field theory and the 
renormalisation group to understand the clustering
behaviour of a self-gravitating expanding universe. We also consider
the thermodynamics properties of the system, assuming quasi-equilibrium
for the range of scales concerned with the non-linear regime and
virialisation. We find an exact mapping between the  self-gravitating gas 
and a continuous field theory for a
single scalar field with an exponential self-coupling. 
This allows us to use  statistical field theory
and the renormalisation group to determine the scaling behaviour.
The small-scale fractal universe can be considered critical with large density
fluctuations developing at any scale. We derive
the corresponding critical exponents. They are very close
to those measured on galaxy catalogs through statistical
methods based on the average density as function of size; these methods
reveal in particular a fractal dimension $D \approx 1.5-2$ 
(Di Nella et al 1996, Sylos Labini \& Amendola 1996, Sylos Labini et al 1996).
This fractal dimension is strikingly close to that observed for
the interstellar medium or ISM (e.g. Larson 1981, Falgarone et al 1991).
We show in the present paper that the theoretical framework based on 
self-gravity that we have developped for the ISM (de Vega, S\'anchez \& Combes
1996a,b, hereafter dVSC) is also  the dynamical mechanism 
leading to the small scale fractal structure of the universe.
This theory is powerfully predictive without any free parameter. 
It allows to compute the $N$-points density correlations without any extra 
assumption.

We first clarify the definition of the average density we use for
a fractal medium in \S 2, present the dynamical equations of the comoving 
fractal in \S 3, and apply our field-theory approach in \S 4.

\section{Correlation functions and mass density in a fractal}

The use of the two point correlation function $\xi(r)$ widely spread
in galaxy distributions studies, is based on the assumption that the
Universe reaches homogeneity on a scale smaller than the sample size.
It has been shown by Coleman, Pietronero \& Sanders (1988) 
and Coleman \& Pietronero (1992) that such an hypothesis could
perturb significantly the results.
The correlation function is defined as
$$
\xi(r) = \frac{<n(r_i).n(r_i+r)>}{<n>^2} -1
$$
where $n(r)$ is the number density of galaxies, and $<...>$ is the volume
average (over $d^3r_i$). The  length $r_0$ is defined
by $\xi(r_0) = 1$. The function $\xi(r)$ has a power-law behaviour 
of slope $-\gamma$ for $r< r_0$, then it turns down to zero 
rather quickly at the statitistical limit of the sample. This rapid
fall leads to an over-estimate of the small-scale $\gamma$.
Pietronero (1987) introduces the conditional density
$$
\Gamma(r) = \frac{<n(r_i).n(r_i+r)>}{<n>} 
$$
which is the average density around an occupied point.
For a fractal medium, where the mass depends on the size as
$$
M(r) \propto r^D
$$ 
$D$ being the fractal (Haussdorf) dimension, the conditional
density behaves as
$$
\Gamma(r) \propto r^{D-3}
$$
This is exactly the statistical analysis used for the interstellar
clouds, since the ISM astronomers have not  adopted from the start 
any large-scale homogeneity assumption (cf Pfenniger \& Combes 1994).

The fact that for a fractal the correlation $\xi(r)$ can be highly
misleading is readily seen since
$$
\xi(r) = \frac{\Gamma(r)}{<n>} -1
$$
and for a fractal structure the average density of the sample $<n>$ is a 
decreasing function of the sample length scale. In the general use of
$\xi(r)$,  $<n>$ is taken for a constant, and we can see that
$$
D = 3 - \gamma \quad .
$$
If for very small scales,
both $\xi(r)$ and $\Gamma(r)$ have the same power-law behaviour, with the 
same slope $-\gamma$, then the slope appears to steepen for $\xi(r)$
when approaching the  length $r_0$. This explains why
with a correct statistical 
analysis (Di Nella et al 1996, Sylos Labini \& Amendola 1996, 
Sylos Labini et al 1996), the actual $\gamma \approx 1-1.5$ is smaller 
than that obtained using $\xi(r)$. This also explains why the amplitude of
$\xi(r)$ and $r_0$ increases with the sample size, and for clusters as well. 

In the following, we adopt the framework of the fractal medium that 
we used for the ISM (dVSC), and will not 
consider any longer $\xi(r)$.

\section{Equations in the comoving frame}

Let us consider the universe in expansion with the characteristic
scale factor $a(t)$. For the sake of simplicity, we modelise the galaxies
by points of equal masses $m$, although they have a mass spectrum
(it may be responsible for a multi-fractal structure, see Sylos Labini
\& Pietronero 1996). 

The present analysis  can
be generalised to galaxies of different  masses  following the
lines of sec. IV in de Vega, S\'anchez \& Combes (1996b). We expect to
come to this point in future work.

If the physical coordinates of the particles are  $ {\vec r} $, we can 
introduce the comoving coordinates $ {\vec x} $ such that 
$$
 {\vec r} =  a(t) \; {\vec x} 
$$
The Lagrangian for a system of $ N $ particles interacting only
by their self-gravity can be written as
\begin{equation}\label{lagra}
L_N = \sum_{i=1}^N \left[ \frac{m}2  a(t)^2 \;
 {\dot  {\vec x}}_i^2 - \frac{m}{
a(t)} \; \phi( {\vec x}_i(t)) \right] \; ,
\end{equation}
where  $ \phi( {\vec x}) $ is the gravitational potential in the
comoving frame, determined by the Poisson equation
\begin{equation}\label{poi}
\nabla^2 \phi( {\vec x})= 4\pi G \; \rho( {\vec x}, t) \; ,
\end{equation}
and $  \rho( {\vec x}, t) $ is the mass density. 
For our system of point particles,
\begin{equation}\label{rho}
 \rho( {\vec x}, t) = m \, \sum_{i=1}^N \delta( {\vec x}-  {\vec x}_i(t))
\end{equation}
and therefore the solution of the Poisson equation takes the form
\begin{equation}\label{fi}
 \phi( {\vec x}) = - G m  \sum_{i=1}^N { 1 \over { |  {\vec x}-  {\vec
 x}_i(t)| }}\; .
\end{equation}

The canonical momenta and Hamiltonian of the system are
$$
{\vec p_i} = m\;   a(t)^2  \; {\dot  {\vec x}}_i 
$$
$$
H_N =  \sum_{i=1}^N \left[{ 1 \over { 2 m a(t)^2}}\;{\vec p_i\,}^2 + \frac{m}{
a(t)} \; \phi( {\vec x}_i(t)) \right] 
$$
\begin{equation}\label{hamC}
= { 1 \over { 2 m  a(t)^2}}\; \sum_{i=1}^N {\vec p_i\, }^2 -  {{G \, m^2}
\over {a(t)}} \sum_{1\leq l < j\leq N} {1 \over { |{\vec x}_l - {\vec x}_j|}} 
\end{equation}
We see that the $N$-particle Hamiltonian in cosmological spacetime
eq.(\ref{hamC}) can be obtained from the Minkowski Hamiltonian
[$ a(t)= 1 $] by making the replacements 
\begin{equation}\label{cambios}
m \to m \, a(t)^2 \quad ,  \quad G \to G \,  a(t)^{-5} \; . 
\end{equation}

As a first approximation, we shall assume in the following that   
the characteristic time of the particle motions under the
gravitational self-interaction are shorter than the
time variation of $a(t)$. We can then consider that this
system of self-gravitating particles is at any time 
in approximate thermal equilibrium.
This hypothesis is true of course for structures that have already 
decoupled from the expansion, and are truly self-gravitating and virialised.
 It could be also valid for the whole non-linear regime of
the gravitational collapse. 
As for the linear regime, we know already that the primordial fluctuations
are not forgotten in the large-scale structures, and therefore the
resulting correlations will depend on initial conditions, and not be entirely
determined by self-gravity.

 The above assumption introduces a natural upper limit in the 
scales concerned by the theory developped below. Similarly to the
case of the interstellar medium, the fractal structure considered is
bounded by a short distance cut-off and by a large-scale limit as well
(dVSC).

 The short distance cut-off corresponds to the appearence of other
physics at  short scale, essentially dissipative, which we do not need
to introduce. In addition, the short distance cut-off
avoids the gravo-thermal catastrophe. For the ISM, 
the cut-off was naturally the size of the smaller fragments,
of the order of the Jeans length. Here the cut-off
corresponds also to the size of the `particles' considered, i.e. the
galaxy size, below which another physics steps in, related to
stellar formation and radiation. 

The present treatment can be  generalized when thermal equilibrium
only holds region by region (de Vega, S\'anchez \& Combes, in preparation).
In such case we are lead to a quenched average over the temperature and we 
argue that the scaling properties are the same as in exact thermal
equilibrium provided the temperature variations are smooth over the structure
at the considered scale.

The fact that in the catalogs, we are observing in projection large-scale
structures at different epochs, with different values of the scale
factor $a(t)$, could slightly modify the fractal dimension. Even though fractal
structures are self-similar, and scale-independent, the largest scales
are systematically observed at a younger epoch where the contrast has not
grown up as high as today. This evolution effect however should be significant
only at high redshift ($>1$), and the present catalogs are not yet
statistically robust so far back in time (the average redshift of optical
catalogs is about 0.1).

\section{Application of renormalization group theory}

As in all scale-independent problems, where the fluctuations cannot
be represented by analytical functions, the renormalization group
theory developped in the 1970's for the study of critical phenomena, appears
here perfectly adapted (e.g. Wilson \& Kogut 1974). We can consider the fractal
structure of the Universe as the critical state of a system, 
where fluctuations develop at any scale, with a very
large correlation length (asymptotically infinite). The fluctuations that are 
distributed as a fractal of dimension $D$ are the large-scale structures of 
the universe (cf Totsuji \& Kihara 1969).

 We have recently begun to tackle, with the tools of statistical field
theory, the study of an $N$-body system only interacting through their
own self-gravity (dVSC).
We have found an exact mathematical connection between this statistical
system and a scalar field with exponential self-coupling. 
We then used for it the powerful methods of field theory 
(e.g. Itzykson \& Drouffe 1989, Parisi 1988, Zinn-Justin 1989). Using the
renormalisation group, the critical 
behaviour of this gravitational system has been described and its
critical exponents identified.
This theory explains both
the origin of the fractal structure, and predicts its fractal dimension $D$.
This has been successfully applied to the interstellar medium (dVSC).
 Another  approach has been proposed for galaxy
correlations (not for ISM) (Hochberg \& P\'erez Mercader
1996), but it yields different critical exponents.

Let us apply the theory to the system of galaxy points, already
defined in the previous section. Since they are considered 
in approximate thermal equilibrium, we will
use the grand canonical ensemble, that also
allows a variable number of particles.
The grand partition function of the system can be written as

\begin{equation}\label{gfp}
{\cal Z} = \sum_{N=0}^{\infty}\; {{z^N}\over{N!}}\; \int\ldots \int
\prod_{l=1}^N\;{{d^3p_l\, d^3x_l}\over{(2\pi)^3}}\; e^{- \beta H_N}
\end{equation}
where $z$ is the fugacity, which  for an ideal gas of number density $\rho_0$
is : 
$z = \rho_0 \; \left({{h^2}\over{2\pi m kT}}\right)^{3/2}$ ($h$ is the Planck constant).

In dVSC we found a functional integral representation for
the grand partition function 
\begin{equation}\label{zfiM}
{\cal Z} = \int\int\;  {\cal D}\phi\;  e^{ -S[\phi(.)] }
\end{equation}
i.e., ${\cal Z}$ can be written as the partition function for a single
scalar field $\phi({\vec x})$  with  {\bf local} action
\begin{equation}\label{acciM}
S[\phi(.)] \equiv  { 1 \over{T_{eff}}}\;
\int d^3x \left[ \frac12(\nabla\phi)^2 \; - \mu^2   \; e^{\phi({\vec
x})}\right] \; . 
\end{equation}
where
\begin{equation}\label{muyT}
\mu^2 = {\pi^{5/2}\over {h^3}}\; z\; G \, (2m)^{7/2} \, \sqrt{kT} , 
\quad T_{eff} = 4\pi \; {{G\; m^2}\over {kT}} 
\end{equation}

In the $\phi$-field representation, the mass density  eq.(\ref{rho}) is
expressed as
\begin{equation}\label{denfi}
  \rho({\vec x}) =  -{m \over {T_{eff}}}\;\nabla^2 \phi({\vec r})=
{{m \, \mu^2}\over{T_{eff}}} \; e^{\phi({\vec r})}  \; .
\end{equation}
and the mass contained in a region of size $ R $ is
\begin{equation}\label{defM}
M(R) = {{m \, \mu^2}\over{T_{eff}}} \int^R e^{\phi({\vec x})} \; d^3x  \; .
\end{equation}
The mass parameter $\mu $ coincides at the tree level with the inverse of the
Jeans length $d_J$ (dVSC)
\begin{equation}\label{longJ}
\mu =  \sqrt{12 \over {\pi}}\; { 1 \over {d_J}} \; .
\end{equation}

The functional representation for the  grand partition function 
can be easily generalized for an arbitrary scale
factor $a(t)$. After the changes specified above in
eq.(\ref{cambios}), the local action becomes
\begin{equation}\label{acci}
S[\phi(.)] \equiv  { {a(t)} \over{T_{eff}}}
\int d^3x \left[ \frac12(\nabla\phi)^2  - \mu^2   a(t)^2 \;  e^{\phi({\vec
x})}\right] 
\end{equation}
Notice that all quantities depend on time
through the scale factor $ a(t) $ only. There is no integration over $ t $. 

The mass parameter $ \mu $  in the $\phi$-theory gets effectively
multiplied by the scale factor  $ a(t) $.  Since the Jeans length $ d_J
\simeq \mu^{-1} $ according to eq.(\ref{longJ}), 
in comoving coordinates $ d_J $ effectively becomes 
$$ 
d_J  =  \sqrt{12\over {\pi}}\; { 1 \over {\mu \, a(t) }} \quad ,
$$ 
as one could have expected.  

On the other hand, the dimensionless coupling constant 
$$
g^2 = \mu \,  T_{eff}
$$
is unchanged by the  replacements of eq.(\ref{cambios}).

Therefore, for any fixed time $ t $  we find the same scaling
behaviour, after making the replacement
$$
\mu \to \mu \; a(t) 
$$
and keeping the coupling $ g $ unchanged. 

\subsection{Scaling behaviour}

As is well known in the theory of critical phenomena 
(e.g. Wilson 1975, 1983; Domb \& Green 1976), physical quantities  for {\bf
infinite} volume systems diverge at the critical point  as $ \Lambda $ to a
negative power, where  $ \Lambda $ measures the distance to the critical
point. The correlation length  $ \xi_0 $ diverges as
$$ 
\xi_0( \Lambda ) \sim  \Lambda^{-\nu} \; ,
$$
and the specific heat $ {\cal C} $ behaves as
\begin{equation}\label{calor}
 {\cal C} \sim  \Lambda^{-\alpha}  \; .
\end{equation}
The critical exponents $\nu$ and $\alpha$ are pure numbers
that depend only on the universality class of the
problem considered (e.g. Binney et al 1992).

For a {\bf finite} volume system, all physical quantities are {\bf
finite} at the critical point. And for a  system whose size $ R $
is large, the  physical quantities take large, but finite, values at 
the critical point. Thus, for large critical systems, one can use 
asymptotically the infinite volume theory.
In particular, the correlation length  $ \xi_0 $ can be identified with
the relevant physical scale $ R  : \;  \xi_0 \sim R $. This implies that
\begin{equation}\label{fss}
\Lambda \sim R^{-1/\nu} \; .
\end{equation}
These concepts apply to the gravitational $\phi$-theory since it 
exhibits scaling in a finite volume $\sim R^3 $ (dVSC).
Scaling behaviour was found for a {\bf continuum set} of
values of $\mu^2$ and $ T_{eff} $.

We have previously shown (dVSC) that it is possible to identify
\begin{equation}\label{zcritico}
  \Lambda \equiv   {{\mu^2}\over{T_{eff}}} = { {z} \over {h^3}}
  \left(2\pi mkT\right)^{3/2} \; .
\end{equation}
Notice that the critical point $  \Lambda = 0 $, corresponds to zero
fugacity. 
Then, the partition function in the scaling regime can be written as 
\begin{equation}\label{Zsca1}
{\cal Z}(\Lambda) = 
 \int\int\;  {\cal D}\phi\;  e^{ -S^* + \Lambda
\int d^3x  \; e^{\phi({\vec x})}\;}\; ,
\end{equation}
where $S^*$ stands for the action  at the critical point.

We define the renormalized mass  density  as
\begin{equation}\label {dfensi}
m\, \rho({\vec x})_{ren} \equiv m\, \,  e^{\phi({\vec x})}
\end{equation}
and we identify it with the  energy density in the renormalization
group (also called the `thermal perturbation operator').

Since the $\phi$-theory exhibits scaling (dVSC), the non-analytical part
of the free energy is
$$ 
\log{\cal Z}(\Lambda) \propto \Lambda^{2-\alpha} \quad ,
$$
so that its second derivative is ${\cal C} \sim  \Lambda^{-\alpha}$.
Calculating the logarithmic derivative of ${\cal Z}(\Lambda)$ with
respect to $ \Lambda $ from eqs.(\ref{Zsca1}),
using the standard relation between critical exponents
in a three dimensional space $\alpha = 2 - 3 \nu $,
and eqs. (\ref{dfensi}) and (\ref{defM}), we find
that the mass fluctuations inside a volume of radius $ R $
$$
(\Delta M(R))^2 \equiv  \; <M^2> -<M>^2  \quad ,
$$
will scale as
\begin{equation}\label{Msca}
\Delta M(R)  \sim  R^{\frac1{\nu}}\; .
\end{equation}

The scaling exponent $ \nu $ can be identified with the inverse
Haussdorf (fractal) dimension $D$ of the system
$$
D = {1\over{\nu}} \; .
$$

\subsection{Critical exponents}

As usual in the theory of critical phenomena, there are only two
independent critical exponents. All exponents can be expressed in
terms of two of them: for instance the fractal dimension $ D = 1/ \nu $,
and the independent exponent $\eta$, 
which usually governs the spin-spin correlation
functions. The exponent  $\eta$ appears here in the $\phi$-field 
correlator (dVSC), describing the gravitational potential, that scales as
$$
<\phi({\vec r})> \; \sim r^{-\frac12(1+\eta)}
$$
The values of the critical exponents 
depend on the fixed point that governs the long range
behaviour of the system. 

The renormalization group approach applied to a
single  component scalar field in three space dimensions
shows the presence of only two fixed points: the mean field point and
the Ising fixed point.
The scaling exponents associated to the Ising fixed point are
$\nu_{Ising} = 0.631...$, $D_{Ising} = 1.585...$ , and 
$\eta_{Ising}=0.037...$ .
 The mean field value for the critical exponents are $ \nu_{mean f} 
= \frac{1}{2} $ , $ D_{mean f} = 2 $ and $ \eta_{mean f} = 0 $.

 The value of the dimensionless coupling constant $g^2 = \mu T_{eff}$
should decide whether the fixed point chosen by the system is the
mean field (weak coupling) or the Ising one (strong coupling). 
At the tree level, we
estimate  $g \approx \frac{5}{\sqrt{N}}$, where 
$N$ is the number of points in a Jeans volume $d_J^3$. The coupling
constant appears then of the order of 1,
and we cannot settle this question
without effective computations of the renormalisation group equations.
At this point, the predicted fractal dimension $D$ should be between
1.585 and 2.

\subsection{Three point and higher correlations}

Our approach allows to compute higher order correlators without any extra 
assumption (de Vega, S\'anchez \& Combes, in preparation).

The two and three point densities,
\begin{eqnarray}
D({\vec r}_1,{\vec r}_2) &\equiv& <n({\vec r}_1)\, n({\vec r}_2)> \cr \cr
D({\vec r}_1,{\vec r}_2,{\vec r}_3) &\equiv& <n({\vec r}_1)\, n({\vec r}_2) \,
n({\vec r}_3)> \; ,
\end{eqnarray}
can be expressed as follows in terms of the correlation functions:
\begin{equation}\label{distr3}
D({\vec r}_1,{\vec r}_2)  =  n_1 \;  n_2 +  C_{12}
\end{equation}
$$
D({\vec r}_1,{\vec r}_2,{\vec r}_3) = n_1 \;  n_2  \;  n_3 + 
 n_1 \; C_{23} +  n_3 \; C_{12} +  n_2 \; C_{13} + C_{123}\; .
$$
Here,
$$
n_i \equiv < n({\vec r}_i)> \quad , \; i=1,2,3 \; ,
$$
and $ C_{ij} $ and $ C_{ijk} $ are the two and three point 
correlation functions, respectively,
$$
 C_{ij} \equiv C({\vec r}_i,{\vec r}_j)
$$
$$
 C_{ijk} \equiv C({\vec r}_i,{\vec r}_j,{\vec r}_k)
$$

The behaviour of $ n_i , \;  C_{ij} $ and $ C_{ijk} $ in the scaling regime
follow from the renormalisation group equations at criticality
(de Vega, S\'anchez \& Combes, in preparation). If we
 do not impose homogeneity at all scales, we find,
\begin{eqnarray}\label{rg123}
< n({\vec r})> \simeq A\; r^{D-3} \; , \cr \cr
C({\vec r}_1,{\vec r}_2) \buildrel{r_1 >> r_2 }\over \simeq B\; r_1^{2(D-3)}
\; , \cr \cr
C({\vec r}_1,{\vec r}_2,{\vec r}_3) \buildrel{r_1 >> r_2, r_3 }\over \simeq 
C \;  r_1^{3(D-3)}\; ,
\end{eqnarray}
where $ A, \; B $ and $ C $ are constants and $ D = 1/\nu $.

We can now derive the three point density behaviour when one point, say $ 
{\vec r}_1 $, is far away from the other two. We find from eqs.(\ref{distr3})
and (\ref{rg123}),
\begin{eqnarray}\label{D123}
D({\vec r}_1,{\vec r}_2) \buildrel{r_1 >> r_2 }\over\simeq 
 A \;  r_1^{D-3} \; n_2 \; + B\; r_1^{2(D-3)} \; , \cr \cr
D({\vec r}_1,{\vec r}_2,{\vec r}_3) 
 \buildrel{r_1 >> r_2, r_3 }\over\simeq A \;  r_1^{D-3} \; 
\left( n_2 \; n_3 + C_{23} \right) \cr \cr
+ B\; r_1^{2(D-3)} \; ( n_2 + n_3 ) + C \;  r_1^{3(D-3)}
\end{eqnarray}
Notice that this expression is dominated by the first term since $ D- 3 < 0 $.

Higher point distributions can be treated analogously in our approach. 
We find that the dominant behaviour in the $N$-points density is 
\begin{equation}\label{Ncorre}
C({\vec r}_1,{\vec r}_2,\ldots,{\vec r}_N) 
 \buildrel{r_1 >> r_i, \; 2\leq i \leq N  }\over\sim r_1^{N(D-3)} 
\end{equation}
Notice that when homogeneity is assumed to hold over all scales, the critical 
behaviour of the $N$-point correlation function involves a factor
$ r_1^{(N-1)(D-3)} $, (Itzykson \& Drouffe, 1989). 

Eqs.(\ref{D123}-\ref{Ncorre}) are qualitatively similar, although
 not identical, to the behaviour inferred assuming the factorized hierarchical 
Ansatz (fhA), (Balian \& Schaeffer 1989). That is,
\begin{eqnarray}\label{fhA}
D({\vec r}_1,{\vec r}_2)^{fhA}={\bar n}^2 \left(1+br_{12}^{D-3} \right) 
\cr \cr
\buildrel{r_1 >> r_2 }\over\simeq 
{\bar n}^2 \left(1+r_{1}^{D-3} \right) 
\end{eqnarray}
$$
D({\vec r}_1,{\vec r}_2,{\vec r}_3)^{fhA}={\bar n}^3+
{\bar n}^3 b\left(r_{12}^{D-3}+r_{13}^{D-3}+r_{23}^{D-3}\right)
$$
$$
+{\bar n}^3Q_3 \left[r_{12}^{D-3} r_{13}^{D-3}+ 
r_{12}^{D-3} r_{23}^{D-3}+ r_{13}^{D-3} r_{23}^{D-3} \right] 
$$
$$
\buildrel{r_1>>r_2,r_3}\over\simeq 
{\bar n}^3\left[1+br_{23}^{D-3}+2r_{1}^{D-3} 
(b+Q_3r_{23}^{D-3})+Q_3r_1^{2(D-3)}\right] 
$$
where  $ r_{12} \equiv |{\vec r}_1 - {\vec r}_2 | $ and so on.
$ b $ and $ Q_3 $ are  constants. Notice that in the factorized hierarchical 
Ansatz, the fractal dimension $ D $ is not predicted but it is a free 
parameter.

We see that the dominant behaviours in eqs.(\ref{D123}) and  (\ref{fhA})
are similar in case  the  scaling exponents $ D - 3 $ are the same.

\section{Conclusions}

 The statistical analysis of the most recent galaxy catalogs,
without the assumption of homogeneity at a scale smaller than the
catalog depth, has determined that the universe has a fractal structure
at least up to $\approx 100 h^{-1}$ Mpc (Sylos Labini et al 1996). 
The analysis in terms of conditional density has revealed that
the fractal dimension is between $D$ = 1.5 and 2
(Di Nella et al 1996, Sylos Labini \& Amendola 1996). We apply a
theory that we have developped to explain the fractal structure
of the interstellar medium (dVSC), which has the same dimension $D$.
The physics is based on the self-gravitating interaction of an
ensemble of particles, over scales limited both at short and
large distances. The short-distance cut-off is brought by other
physical processes including dissipation.
The long-range limit is fixed by the expansion time-scale.
In-between, the system is assumed in approximate thermal equilibrium.
  The dynamical range of scales involved in this thermodynamic
quasi-equilibrium is at present limited to 3-4 orders of magnitude,
but will increase with time. 

 The critical exponents found
in the theory do not depend on the conditions at the cut-off, which
determine only the amplitudes. The theory is
based on the statistical study of the gravitational field: it is shown
that the partition function of the N-body ensemble is equivalent to 
the partition function of a single scalar field, with a local action.
This allows to use field theory methods and the renormalisation group
to find the scaling behaviour. We find scaling behaviour for a {\bf
full range} of temperatures and couplings.
The theory then predicts for the system a
fractal dimension $D= 1.585$ for the Ising fixed point,
or $D=2$ in the case of the mean-field fixed point. Both are compatible
with the available observations. The $N$-points density correlators
are predicted to scale with exponent $(N-1)(D-3)$ when $ r_1 >> r_i, \;
2\leq i \leq N $. That is, $ -(N-1) $ for the mean field, or $ -1.415\,(N-1) $
for the Ising point.

We predict in addition a critical exponent
$ - \frac12 (1 + \eta ) $ for the gravitational potential: that is, $ -0.500
$ for mean field or $ - 0.519 $ for the Ising fixed point.

\end{document}